# Deterministic actions on stochastic ensembles of particles can replicate wavelike behaviour of quantum mechanics: does it matter?

Dale R. Hodgson and Vladimir V. Kisil

November 26, 2019

**Abstract**

The probabilistic interpretation of quantum mechanics has been a point of discussion since the earliest days of the theory. The development of quantum technologies transfer these discussions from philosophical interest to practical importance. We propose a synthesis of ideas appeared from the field's founders to modern contextual approaches. The concept is illustrated by a simple numerical experiment imitating photon interference in two beam splitters. This example demonstrates that deterministic physical principles can replicate wavelike probabilistic effects when applied to stochastic ensembles of particles. We also reference other established experimental evidence of the same phenomena. Consequences for quantum information technologies are briefly discussed as well.

## 1 Introduction

The recent quantum technology roadmap [1] invites a wider community to discuss fundamental aspects of the emerging quantum technology. It is instructive to compare the present proposal with an earlier one [24], see also the discussion in [15]. Clearly, we now have a better understanding of the challenges confronting further development and a more cautious prediction of expected progress. Yet some more fundamental discussion may still be relevant.

Superposition of quantum states is a crucial element of quantum information as highlighted in the abstract of [1], see also [4, 39]. However, the meaning of superposition depends on the interpretation of the wavefunction:

- whether it describes the nature of an individual quantum object, a qubit, say; or
- it only exists as a representation of the properties of an ensemble of identical objects.

An interpretation of the wavefunction was disputed since early years of quantum theory, which does not prevent it, however, to be used for accurate experimental predictions. Obviously, taking the wavefunction for granted is only justified for experiments with large ensembles of objects. The emergence of quantum technologies reliant on individual quantum system, e.g. qubit, reignites the significance of the wavefunction question. In this notes we review some classical works on "wave mechanics" and revive their ideas with a new twist.

# 2 Origins of wave mechanics in relativity and paradoxes of measurement

Nowadays, the fields of quantum mechanics and relativity are often treated as at-odds to each other, or as being at least technically incompatible. The nature of entanglement, EPR paradox and Bell's theorem cause the main disagreements between the two theories. Recent experiments [21, 22] claim to fully demonstrate the instantaneous action-at-a-distance effects of quantum mechanics, in violation of the local and continuous model of relativity. However, in the formative years of quantum mechanics, relativity was not seen as such an antagonist to quantum theory. Moreover, in the 1920s relativity was considered as a new fundamental law of nature which can be the ultimate source for any physical theory.

In the celebrate paper [13] de Broglie introduced a wave-like nature of quantum objects considering special relativity laws, see also his later revision [14]. He elegantly combined quantum theory with relativity and lay the foundations for what would become Pilot-Wave Theory. Similarly, but often overlooked now, it is the call to relativistic principles which Schrödinger included in work on the uncertainty principle, see recent discussion in [25]. This confirms that the main mathematical tools of new quantum mechanics—the wave function and the Schrödinger equation—are not fundamentally in odds with relativity.

The disagreements of the two theories (or rather their interpretations) was unfolded later by the famous EPR paradox [16] which relies on the wavefunction designated to describe an individual quantum system. Furthermore,

‘wavefunction collapse’ and ‘projection postulate’ have to be introduced to explain effect of repeated measurement of an individual system. It is not an exaggeration to say that all ‘paradoxes’ of quantum mechanics are rooted in attribution of the wavefunction to an individual quantum system. A good illustration is the textbook [20], where the authors managed to avoid all difficult questions of interpretation until the very last chapter, which introduces wave function collapse and quantum measurement. Attempts to ‘resolve’ their paradoxical consequences open the door to even more exotic theories of many-worlds or even many-minds types [2, 17].

# 3 De Broglie–Bohm pilot wave theory and contextuality

The existing mathematical model of quantum theory gives a good description of physical observations, but does not attempt to explain any mechanism for such results. This omission is raised by John Bell as an argument that current quantum theory cannot be an ultimate one [6]. Other interpretations of quantum mechanics have been proposed, here we choose to focus on those of a contextual nature such as [19, 27, 28].

The use of ‘realism’ in quantum mechanics has been a point of discussion since the days of Bohr and Einstein. Most crucially, the famous Einstein–Podolsky–Rosen paper [16] discussed the place of *elements of physical reality* in quantum theory. In light of formal ‘no-go’ theorems [5, 10], it is usually accepted that realism has no place in a quantum framework. Here, we are inclined to share the view of contextualists such as Khrennikov, that realism *can* be included, provided we accept that any *physically measured result* is a product of both ones target object *and* the system with which it has interacted in order to show such result. Interestingly, the original Niels Bohr’s viewpoint may be more accurately reflected in contextuality rather than by the present dried-out flavour of the orthodox Copenhagen interpretation [28].

Indeed, it should seem natural to consider that any object which never interacts with any other system in the universe has, in effect, no part in reality, and that in any situation where we have learned some physical property of an object, we have done so only by the interaction of said object with some measuring apparatus. Furthermore, it should seem unnatural to take the view of the quantum mechanics orthodoxy, that a ‘measurement operation’ on a quantum system is some totally disparate hand of God which collapses a wavefunction to some result and then returns to the aether. The observation of an experimental outcome is always the result of interaction between object and apparatus, but the apparatus itself must also be a physical body and hence subject to some effect of the interaction. It is this final point, that apparatus is mutually affected by all its quantum interaction, which we shall espouse in our proposed model.

Since the above assumptions are very plausible it is natural that they already appeared in the literature. Different models of such interaction of a quantum system and apparatus were independently considered in [26, 30]. A few years later deterministic objects with wave-like collective behaviour were physical realised as droplets [11]: these are ensembles of well-localised classical objects which interact with each other through waves spreading in the surrounding media. One can easily interpret this framework as an explicit materialisation of de Broglie–Bohm pilot waves theory [8, 12, 23, 38], which were approached from a different theoretical setup in [37]. Closer to the quantum world, there is a recent confirmation of the concept performed on photons [18]. It is not surprising that ideas similar to [26, 30] were independently proposed in the later works, e.g. [34].

In fact the previous discussion [30] can be linked to one of lesser noted work [7] on wavefunctions and we shall adopt key parts of its notation. In this work [7] Bohm and Bub proposed that the probabilistic results of quantum mechanical experiments could be explained by assigning not just a single wavefunction $|\Psi\rangle$ to a quantum objects, but also a second dual-space vector $\langle\xi|$, then with some simple deterministic rules, the outcome of a measurement depends on the states of both vectors together [7].

Our proposal is *to decouple this second vector $\langle\xi|$ from the quantum system and attach it to the measuring apparatus or the contextual environment in a general sense.* In other words, we take Bohm&Bub's dual-vector to represent the quantum state of the physical apparatus which necessarily interacts with a quantum object in order to output a measurement result.

Clearly, this attribution does not change the mathematical theory used in [7] and preserves all desired logical consequences. On the other hand, the proposed merge of the vector $\langle\xi|$ to apparatus will be completely in the spirit of the contextual interpretation of quantum mechanics and will restore the epistemological balance between a quantum system and apparatus during the measurement process. Furthermore, we eliminate a need for a carrier media (a sort of ether) for the pilot wave as well as the necessity to consider an 'empty wave function', which spreads in the vacuum and is not carrying energy or momentum [38]. As a byproduct, the vector $\langle\xi|$ being attached to a physical object—the measuring apparatus—removes a need for non-local theories, cf. [29].

We want to clarify that Bohm and Bub in [7] did not challange the orthodox attribution of the wavefunction to a single quantum system, while our present suggestion is to associate the wavefunction to an ensemble of quantum objects interacting with the same apparatus. Thus, reusing some mathematical technique from [7] we completely revise it interpretation.

# 4 Illustration by a numerical experiment

As a simple illustration of the proposed framework we produce the following numerical experiment for the textbook example of single-photon interference through two beam splitters [4, § 2.A.1], cf. Fig. 1. In this common experiment, a single photon source emits quanta towards a 50/50 beam splitter (BS1), with its two output paths aligned to intersect with a second beam splitter (BS2). The arrangement is such that the path length of one route may be varied. The value $e^{i\delta}$ representing the phase change caused by an increased path length, and the two counters D1 and D2 simply detecting the final position of a photon after passing through the system.

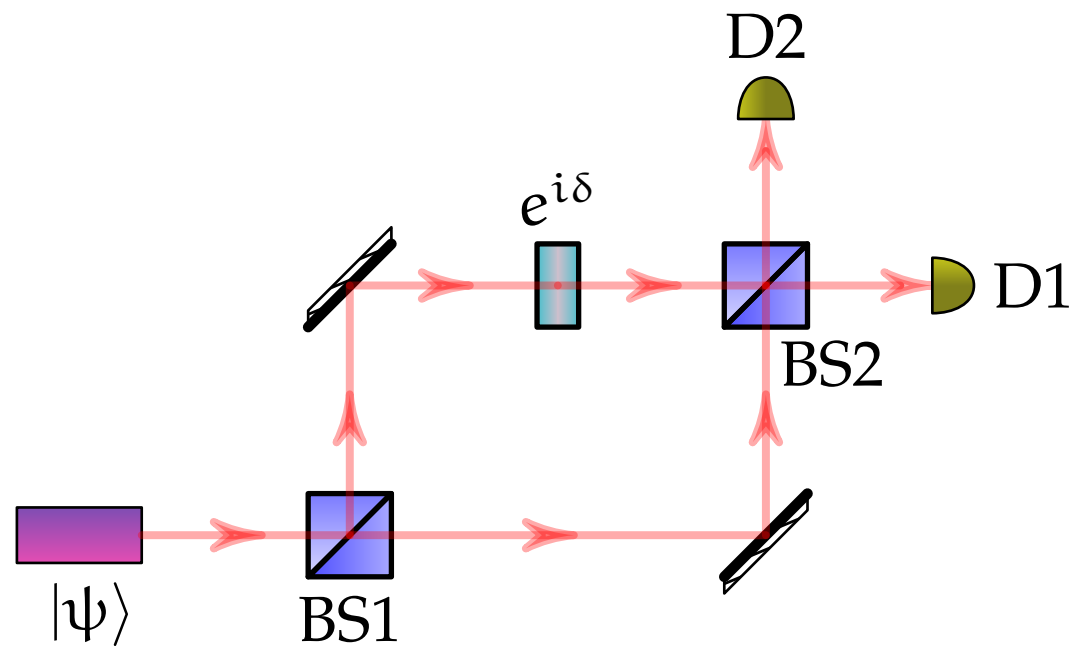

Figure 1: The standard two beam splitters arrangement in Mach-Zehnder interferometer.

There are no claims that our numerical model represents the actual physical setup. Instead we pursue the following two more modest goals:

1. Present an explicit implementation of the proposed two wave functions contextual model.

2. Demonstrate that the new model successfully replicates quantum behaviour through deterministic particles, similarly to early theoretical models [26, 30] and physical experiments with droplets [11] and photons [18].

In a sense, our numerical model is a contemporary variation of 'thoughts experiment' concept used for a long time.

In a wave model, this experiment is simply described by the first beam splitters halving the intensity of incident waves, and the second beam splitter recombining the waves from both paths. When the path length of one route is altered, the waves arrive out of phase and we see wave-like interference of the light by different intensity registered by detectors D1 and D2. A particle model on the other hand, where we may introduce individual photons to such an apparatus, struggles to describe the observed interference effects without introducing a wave-particle duality to our model of light.

We seek to describe a deterministic particle model showing interference effects without the introduction of any dual nature. We do this by considering the '*internal periodic phenomenon*' which formed the beginnings of de Broglie's work on waves and quanta [13, 14]. We model the internal periodic phenomenon as a complex phase with some given frequency $\nu$ and initial value $\phi_0$:

$$\exp\left(i(\nu t + \phi_0)\right).$$

Where the wave model tells us that *two* waves meet and interact at the second beam splitter, we wish to describe each photon as having travelled a single definite path. Then we must allow for the interaction between phases of each particle. To allow for the interference between two particles emitted at different points in time, we note the physical context of the apparatus. In effect, the apparatus has its own internal periodic phenomenon. The interaction of both wave phenomena of the particle and the apparatus changes both by some deterministic rule.

Effectively the wave phenomenon of the apparatus serves as some 'memory' of the phase of particles with which it interacted before. Then, subsequent particles passing the apparatus are affected by their predecessors. This is fully in line with our previous argument that any experimental apparatus is itself a quantum system, and hence must posses its own internal periodic phenomena. Other 'toy models' also exist supporting this variety of theory [30, 34]. Obviously, the interaction of particles with subsequent emissions clearly falls within the realms of locality, giving hope that such models may help to explain more adequately the effects typically attributed to instantaneous wavefunction collapse.

We have a simple numerical simulation of such a model. It is based on the following assumptions:

- Photons are emitted from a coherent source at random time intervals.
- The parts of the apparatus (beam splitters) with which the photons interact have their own phase and frequency, comparable to the $\langle\xi|$ vector of Bohm's 'double solution'.

- The corpuscular photons have an instantaneous interaction effect with beam splitters.

- At the instant of interaction, the beam splitter *reflects* (sending the photon down path 1) if and only if the *phase difference* between the particle and beam splitter is less than $\pi$. Otherwise, the particle is *transmitted* (taking path 2).

- If a reflection occurs, the phases take new values, proportional to the phases at the moment of interaction. If a particle passed through without reflections both phases are unchanged.

In other words, the post-interaction phase change acts as the 'apparatus memory'. Note that if the two paths available are of different lengths, and a particle has taken a path *different* to its predecessor, then there will be a difference between particles' phases and cumulative 'memory' phase of the second beam splitter.

The proposed realisation, see Appendix A, of this model demonstrates the following features:

1. The 50/50 distribution of reflection/transmission seen for a stream of photons interacting with a single beam splitter.

2. A variation from 50/50 distribution for two beam splitters, the achieved maximal deviation is around 25/75.

3. The final reflection rate varying periodically with the alternation of path length between paths 1 and 2.

Indeed, a computer simulation of $10^5$ particles shows a correct distribution in one beam splitter, and over 50 steps in optical path length difference succeeds in showing an interference effect, see Fig. 2.

# 5 Conclusion: why does it matter?

It is a widespread believe that one photon interference observed in Mach-Zehnder interferometer firmly attributes a wave function to an individual quantum object [4]. However, this interpretation requires the 'wave function collapse' to explain results of consecutive measurements. Furthermore, the wave function collapse introduces a variety of paradoxes [20] with action-at-a-distance as an example. The corresponding concept of quantum entanglement [20] is so fragile that its dialectic opposition—spontaneous and uncontrolled disentanglement—is often introduced. An attempt to employ these paradoxes for, say, quantum cryptography [1] puts such future technologies on a potentially shaky basis [15].

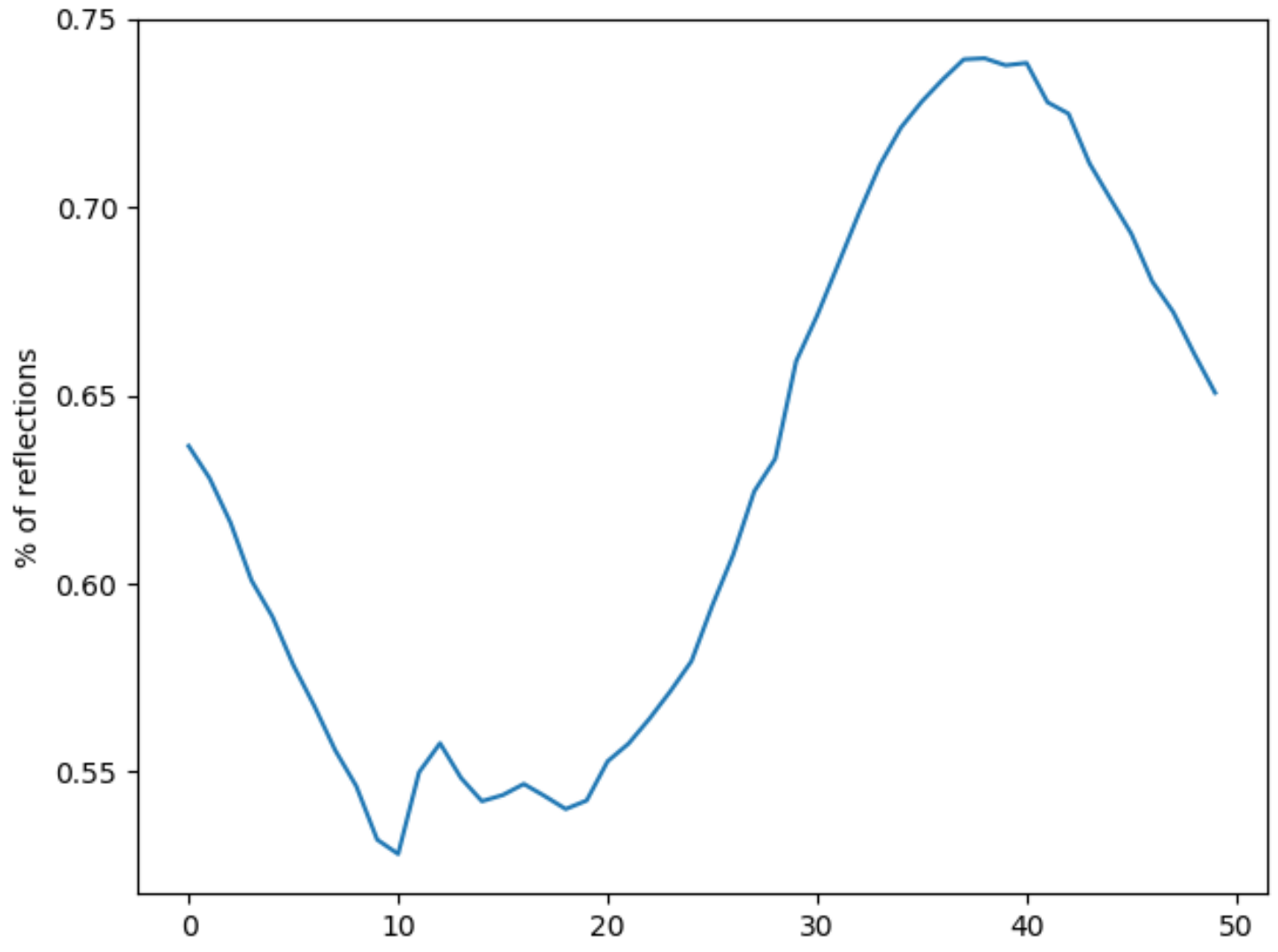

Figure 2: Results of the numerical experiment: there is sine-like dependence of percentage of reflections from the optical paths difference.

The numerical simulation presented in this paper shows that one-photon interference is compatible with an interpretation of the wave function as a description of large ensembles of deterministic particles which self-interact throw the surrounding context. This picture is a blend of the de Broglie–Bohm pilot wave theory [8, 12, 23, 38] with the contextual interpretation of quantum mechanics [19, 27, 28]. This viewpoint is also support by physical evidences [11, 18], however some new specially designed experiments are required to test reality of the de Broglie–Bohm pilot wave and deterministic quantum paths [37]

Combing some classical ideas with recent developments, we attribute the second wave function from [7] to the measuring apparatus or more generally to the context of the quantum system. This does not alter the mathematical model of quantum mechanics and preserves all its predictions for an ensemble of quantum objects described by the traditional wave function. However, consequences for an individual quantum object are significant: they can have deterministic nature as was theoretically predicted [26, 30, 34] and experimentally observed [11, 18].

Possibly, the attribution of the second wave function to the context is rather a convenience and aesthetic choice than a necessity. Similarly, one cannot be forced to accept the Galilean model of the Solar system and may use sufficiently elaborated Ptolemaic epicycles for successful launches of satellites. Yet there are important practical consequences from our discussion. Namely, statistical observations of quantum systems do not prove that an individual trapped ion is caring a superposition of pure states and thus can be used as an implementation of a qubit as it is commonly assumed [1, 24]. Instead, a realisation of a single quibit may requires a management of a large ensemble of quantum objects *and* sufficient control of their self-interaction through the context in order to avoid their disentanglement. Those two tasks are on a completely different technological scale from an operations on a single particular trapped ion [15].

It is a common perception now that the next big step in quantum information requires much more efficient noise control. Optimistically it is compared to tackling static electricity in the early days of lamp computers. However, it is not excluded that 'it is only the noise' problem for quantum computers can be on par with 'it is only the friction' obstacle to create a perpetual movement. In other words, theoretical foundations of quantum information technology require further scrutiny and broad discussion [15, 31].

As a final remark, our work shall not be confused with a search of probabilistic model of quantum mechanics, cf. [3, 9, 35]. We do not try to remove complex-valued amplitudes from the theory and replace them by real (but possibly negative) probabilities. The role of complex scalars in quantisation [32, 33, 36, 37] can be hardly overestimated.

**Acknowledgement.** We are grateful to Prof. A. Khrennikov for useful discussions of this topic and suggestion of some relevant references.

# Appendix A Source code

```python
from math import *
from cmath import *
import matplotlib.pyplot as plt
from random import *

# Initialise the random number generator
seed()

##################################################
# Functions and preliminaries
##################################################

# Phases are normalised to the interval [0,2]
# for simplicity. Variable phase for beam
# splitter 1 (BS1), Xi1, starts from
# a random value.
Xi1 = 2.0*random()

# Variable phase for BS2, Xi2, starts from
# a random value.
Xi2 = 2.0*random()

# Measure of how beamspliter atom is heavier
# than photon.
ratio = 20.0
```

```
# Function describing the effect of a beam splitter.
def BeamSplitter(Phase, Xi):

    # Difference between photon phase (Phase), and beam
    # splitter phase (Xi).
    Delta = (Phase-Xi) % 2
    reflection = ( abs(Delta-1.0)<.5)

    if (reflection): # Reflection condition.

        # Both phases are transformed by the interaction:

        # Beam splitter phase is changed.
        Xi = (Xi+Delta/(1.0+ratio)) % 2

        # Photon phase is changed, with a .5 offset.
        Phase = (Phase+.5-Delta*ratio/(1.0+ratio)) % 2

    return Phase, Xi, reflection

# Function to check the distribution in a single beam
# splitter.
def PhotonInOneBS(Phase):

    global Xi1

    _, Xi1, reflection=BeamSplitter(Phase, Xi1)

    if (reflection):
        return 1
    else:
        return 0

# Function to simulate two consecutive beam splitters by
# running the BeamSplitter() procedure twice, introducing
# an optical difference if the (simulated) photon is
# reflected by the first beam splitter.
def PhotonInTwoBS(Phase, OpticalDifference):

    global Xi1
    global Xi2

```

```python
    # First beam splitter.
    Phase, Xi1, reflection = BeamSplitter(Phase, Xi1)

    # Add the optical path difference for reflected photon.
    if (reflection):
        Phase = (Phase+OpticalDifference) % 2

    # The same rules are applied to the second beam
    # splitter but the outgoing phase is of no consequence.
    _, Xi2, reflection = BeamSplitter(Phase, Xi2)

    if (reflection):
        return 1
    else:
        return 0

##################################################
# Simulated experiment
##################################################

# Number of photons in each experiment.
N = 100000

# Count of reflections.
Ref=0.0

# Using PhotonInOneBS() to check the distribution of
# single photons under these rules. Simulate a sequence
# of N photons interacting with a single beam splitter:
for i in  range(N):

    # Count of photons leaving ‘reflection’ output
    # of beam splitter.
    Ref = Ref+PhotonInOneBS(2.0*random())

print("Percent of reflections in one beam splitter", \
      1.0*Ref/N)

# Using PhotonInTwoBS() to simulate the two beam splitter
# experiment for a range of optical path differences.

# Subdivisions of the optical path difference.
Steps = 50
```

```

# Array of results.
Gr=[]

#Run through all possible optical path differences:
for j in  range(Steps):

    # Count of reflections.
    Ref=0.0

    # Randomly re-initialise phase for BS1 Xi1.
    Xi1 = 2.0*random()

    # Randomly re-initialise phase for BS2 Xi2.
    Xi2 = 2.0*random()

    # Simulate a sequence of N photons:
    for i in  range(N):

        # Count of photons leaving given output of
        # second beam splitter.
        Ref = Ref+PhotonInTwoBS(2.0*random(), \
                                      2.0*j/Steps)

    # Log reflection rate for given optical path\
    # difference.
    Gr.append(Ref/N)

    print(j, Ref/N)

# Plotting the results of two beam splitter simulation.
plt.plot(Gr)
plt.ylabel('% of reflections')
plt.show()
```